# Creating an extrovert robotic assistant via IoT networking devices

Panagiotis Doxopoulos,
Konstantinos L. Panayiotou,
Emmanouil G. Tsardoulias,
Andreas L. Symeonidis

School of Electrical and Computer Engineering
Aristotle University of Thessaloniki - AUTH
Thessaloniki 54124, Greece

*Abstract*—The communication and collaboration of Cyber-Physical Systems, including machines and robots, among themselves and with humans, is expected to attract researchers' interest for the years to come. A key element of the new revolution is the Internet of Things (IoT). IoT infrastructures enable communication between different connected devices using internet protocols. The integration of robots in an IoT platform can improve robot capabilities by providing access to other devices and resources. In this paper we present an IoT-enabled application including a NAO robot which can communicate through an IoT platform with a reflex measurement system and a hardware node that provides robotics-oriented services in the form of RESTful web services. An activity reminder application is also included, illustrating the extension capabilities of the system.

## I. INTRODUCTION

Internet of Things (IoT) is in a boosting period with recent reports showing that by 2020 the connected devices will reach the number of 50 billion [1]. Along with IoT, another technological field that is expected to play a major role in tomorrow's society by helping peoples' daily activities is robotics [2]. In fact, robotic applications are developed in the same domains as IoT technologies: industrial, smart city, health-care and military are some of the most typical domains [2], [3]. The coupling of these two technology fields could offer additional capabilities, as an IoT platform could provide the connected robots with access to smart devices or the ability to use external web services. This way more sophisticated applications may be designed and built.

Taking the above into consideration, this paper aspires to demonstrate the advantages of the coupling of technologies. Specifically, the paper presents a system where an IoT platform is the central component that allows the interaction between humans, robots, things and systems. A hardware node providing robotic web services has been implemented and connected to the platform. Moreover, the IoT platform provides connection to a medical system that can accurately measure reflexes.

In our application, we have used NAO, which is an autonomous, programmable humanoid robot developed by Aldebaran Robotics. It is probably the most widespread humanoid robot, especially in the academic and scientific fields. NAO has 25 degrees of freedom, offers connectivity with Ethernet and WiFi, can be programmed with several programming languages and has a significant number of sensors and actuators.

## II. ARCHITECTURE & IMPLEMENTATION

In this section, the application's architecture is presented, separately describing each component. Fig. 1 shows an overview of the connected components to our platform. The system consists of a router, a hardware node that provides local or remote RESTful robotic web services (REST server), a NAO robot following a specific architecture (R4A[1]) and a system for measuring reflexes (REMEDES [2]). Users can directly interact with the robot and the REMEDES system. Moreover, the platform allows clients, such as doctors, to connect to the platform and get various measurements using the pub/sub protocol.

### A. Crossbar Router

An IoT platform is defined as the middleware and the infrastructure that enables the end-users to interact with smart objects [4]. Nowadays, there is an abundance of middleware solutions. Each company develops its own platform, according to the requirements and the needs of its customers. Crossbar.io [3] is an open source networking platform for distributed and microservice applications. It supports Web Application Messaging Protocol [4] (WAMP) and REST architectural styles[5], through REST bridging. WAMP is an open standard Web Socket subprotocol that is useful for IoT applications, while REST defines a set of architectural principles by which Web services that focus on a system's resources can be designed, including how resource states are addressed and transferred over HTTP [5]. Communication is accomplished using Remote Procedure Calls (RPCs) and Publish/Subscribe (PubSub) protocols. Crossbar.io is an advanced WAMP router made and supported by Crossbar.io Gmbh (creators of WAMP).

---
[1] http://r4a.issel.ee.auth.gr
[2] http://remedes.eu
[3] https://crossbar.io/
[4] http://wamp-proto.org/
[5] https://www.ics.uci.edu/~fielding/pubs/dissertation/rest_arch_style.htm

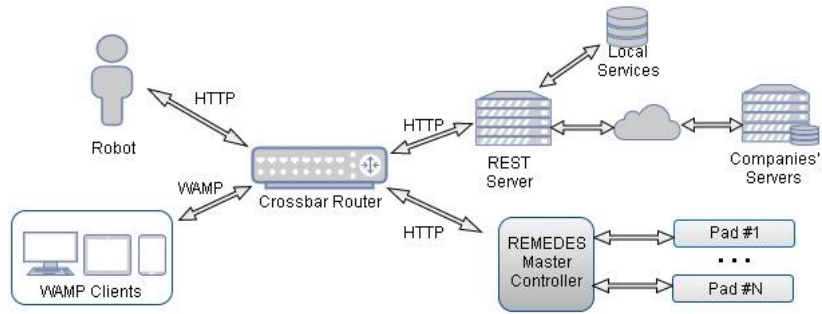

*Fig. 1 Overview of system*

Crossbar router is a core component of the system, as every other component is connected to it, forming this way a star network topology. For the needs of this paper, Crossbar.io router was installed in a laptop's python virtual environment. Properly designing configuration files allows the communication, REST or WAMP, between the connected devices. Crossbar.io router can be fully configured from a single configuration file in YAML format[6].

### B. RPI3 Hardware Node providing robotic web services

There is an abundance of embedded IoT devices, nevertheless Raspberry Pi stands out because of several factors like cost, power consumption, on-board connectivity modules (WiFi and BLE), extended software support, and more importantly a huge community and practitioners with years of experience. R-PI is a credit card sized single board computer weighting only 45g. Raspberry Pi 3 Model B has a quad core 1.2GHz processor and 1GB program memory (RAM). Therefore, it is low cost, powerful, it does not consume a lot of power and it has Wi-Fi connectivity which makes it perfect for integration to an IoT system. Several applications have been performed using R-PI devices, ranging from home automations [6] to monitoring electro-cardiogram signals [7].

Regarding the web server implementation, we used the most popular framework of developing RESTful APIs, Swagger[7]. We preferred Swagger over RAML and API Blueprint, since it has the largest and most active developer community [8] and also provides useful tools for describing, producing, consuming and visualizing web services. The most important tools are Swagger Editor, Swagger Codegen and Swagger UI, while OpenAPI specification is the definition format and schema that describes the APIs. It is also important to mention that Swagger can autogenerate Server and Client code in a plethora of programming languages, thus no need for manually developing these modules existed.

In our architecture, an RPI3 node hosts a server that was designed and built following OpenAPI specification and by using Swagger tools, server and client stubs were generated. The server provides the ability to call robotic services via REST over HTTP. The services are intended for use by robots and are either locally installed (on Raspberry Pi) or are invoked by proxy servers. RAPP[8] services are installed locally while proxy Mashape[9], Algorihmia[10] and Angus[11] services can be called via the world-wide web. RAPP was an FP7 research project, one of whose outcomes was a Cloud-based platform which hosts robotic services, named RAPP Platform.

### C. R4A Architecture

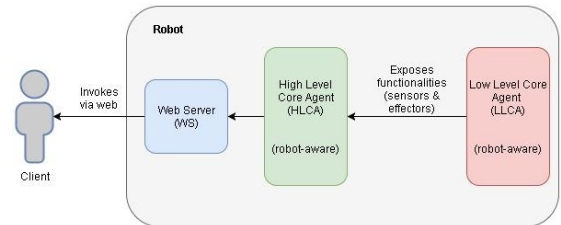

*Fig. 2 The R4A architecture*

A NAO robot was connected to the IoT platform. The NAO robot's middleware follows the R4A architecture (created by R4A group) through which remote or non-remote control of robot functions is provided. This architecture consists of three levels (LLCA, HLCA and Web Server) as shown in Fig.2. Low Level Core Agent (LLCA) is a fundamental part of the R4A architecture comprising all the necessary functionalities to expose the robot resources (actuators and effectors). For utilization of robot resources, this level includes all the necessary drivers and software. High Level Core Agent (HLCA) directly interacts with the LLCA in a robot-aware manner in order to expose hardware calls to the outer world, while the Web Server (WS) module uptakes the task of launching an in-robot server listening for requests. WS module is auto-generated using the swagger-codegen tool[12] and allows synchronous, asynchronous and event-driven communications. The NAO robot was mainly used as the proxy between the system and the user in terms of communication, but also as a humanoid entity the user can be comfortable with, in comparison to a PC or tablet.

---

[6] http://yaml.org/
[7] https://swagger.io/
[8] http://rapp-project.eu/
[9] https://www.mashape.com/
[10] https://algorithmia.com/
[11] https://www.angus.ai/
[12] https://github.com/swagger-api/swagger-codegen

### D. REMEDES system

The REMEDES system can accurately measure reflexes and it is suitable for medical (e.g. diagnosis, neurokinetic) and sports applications (e.g. modeling the reaction of athletes). The architecture of REMEDES is depicted in fig. 3.

Each Pad consists of a Raspberry Pi 3 and a custom-made PiHAT that incorporates four LEDs and a sonar sensor (fig. 4). It also has software controllers for controlling the LEDs and the sonar (take measurements). In addition, each pad provides REST web services, such as the activation service, which triggers the pad by having the trigger time, color, distance and more as parameters. As for the Master Controller, it is the device that controls the Pads. In fact, this is another Raspberry Pi 3 that features: (i) an API, the Pad Client, for communicating with the Pads, (ii) a controller that uses the API to handle the Pads, and (iii) a web server that provides a web-based graphical user interface (GUI) through which the controller functions can be called. Moreover, there is a cloud infrastructure that backs-up all the data and allows users to manage their accounts. The users can get a detailed report of results using the front-end of the cloud platform (Cloud UI).

A standard REMEDES routine involves the sequential activation of Pads providing either visual or audio stimuli, where the user has to deactivate them by moving (usually) their hand in a predefined distance from the Pad. Then, the sonar sensor detects this motion, deactivates the Pad and the Master controller activates the next Pad.

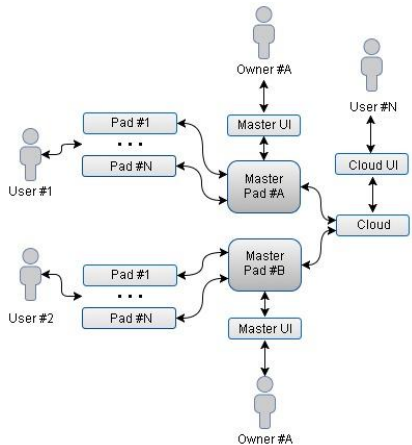

*Fig. 3 The REMEDES system*

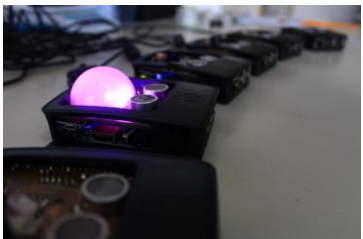

*Fig. 4 The REMEDES Pads*

## III. APPLICATION DESCRIPTION

### A. Description

The aim of the developed application is to remind users about their activities including the REMEDES exercise. Useful data like date, time, activities and activities' recursions are stored in an SQL database (MySQL), thus being treated as a calendar. The overall application consists of two smaller apps that are executed independently of each other.

The first application (fig. 5) allows a user to store an activity to the database. The user interacts with the robot in order to provide information about each activity and its timing. The robot records user's answer and sends the sound file to the application. Then, the audio file is sent through the HW node to the RAPP service *textToSpeech*, which converts audio to text. Next, the text (user's answer) is sent to another service through the HW node. The **Reminders and Events NLP** Mashape service that extracts information, such as date, time and activity, from the text is being used[13]. All the necessary information is finally stored in the database.

The second application (Fig. 6) is executed repeatedly aiming at informing the user of the time to perform an activity. If the time and date of stored activities coincide with the real time and date respectively, then the robot verbally informs the user of which activity he/she must perform. Moreover, the application gives the ability to perform REMEDES exercises. If it's time to execute the exercise, the robot informs the user and the exercise starts automatically, initiated by the robot. When the exercise is done, NAO retrieves the results and informs the user about their achievements. Concurrently, the router publishes the results to a topic using PubSub over WAMP protocol. Doctors and other users can subscribe on the same topic in order to retrieve the results.

### B. Used endpoints

This section provides a brief description of the services that were requested, using HTTP verbs, during the execution of the application.

- *POST* **NAO::Speak**: The RAPP robot API provides an interface to a speech synthesis module. We use this API call to directly invoke speech synthesis in order for NAO to speak.
- *GET* **NAO::Record**: Record user's speech using NAO microphones.
- *POST* **RPi::SpeechRecognition**: We use the RAPP platform speech recognition service provided by the HW node (RPi). Sending the audio file to the service returns the words or phrases contained therein.
- *POST* **RPi::Reminder**: Mashape service to extract useful information from a phrase. The service is providing through the HW node and returns information such as date, time and activity. If the activity is repeated it also returns the recursion time.
- *POST* **REMEDES::startExercise**: Making a request to REMEDES system with predetermined values so as to start the exercise.

---

[13] https://market.mashape.com/maciejgorny/reminders-and-events-nlp

- *GET* **REMEDES::getResults**: REMEDES returns the results of the exercise.

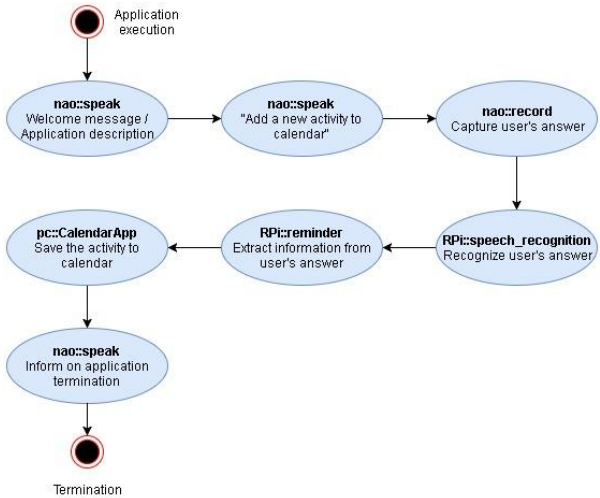

*Fig. 5 Calendar Application: Storing activities to the database*

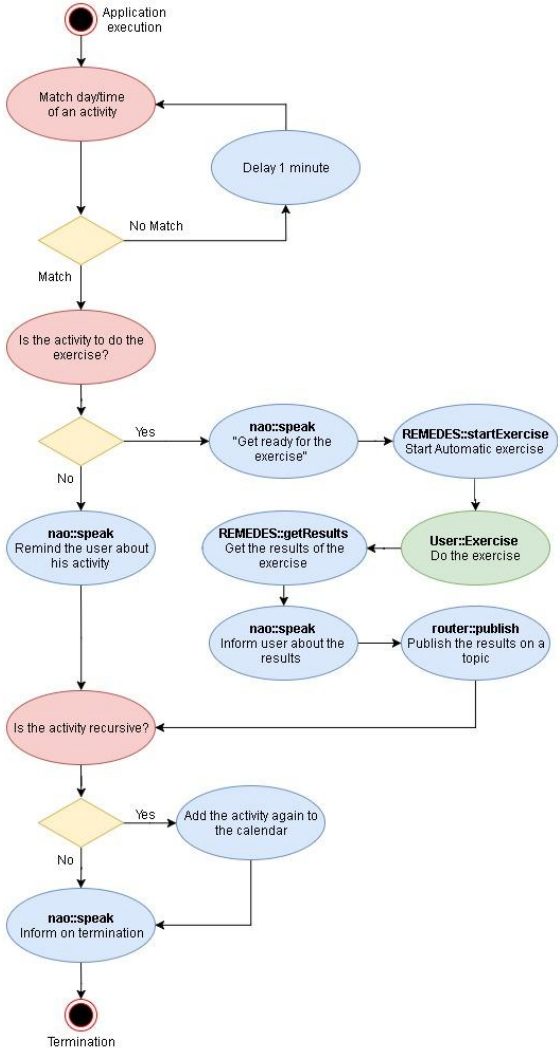

*Fig. 6 Reminder Application: Informs the user about an activity*

## C. Application example

For example, the user may want to get informed when it's time for his/her medicine or set a date to exercise with REMEDES. The user dictates to NAO the phrases: "*Remind me to take the medicine every day after lunch. Furthermore, remind me to practice REMEDES on Sundays nights*". The used Mashape service returns the values: *{year: 2017, **month**: 10, **day**: 18, **hour**: 14, **minute**: 0, **body**: take the medicine, **recurring**: yes, **repeat**: DAYS, 1}* and *{year: 2017, **month**: 10, **day**: 22, **hour**: 20, **minute**: 0, **body**: practice REMEDES, **recurring**: yes, **repeat**: DAYS, 7}*. Then, results are stored in the database and when the time comes, the robot informs the user by saying "*Remember, you must take the medicine*" or "*It's time to practice REMEDES!*". Finally, the application checks if the activity is recurring and adds a new entry to the database with parameter "**day**" equal to "19" for the medicine or "**day**" equal to "29" (next Sunday) for the REMEDES system. Then the process is repeated.

## IV. CONCLUTION & FUTURE WORK

In this paper, we have presented a system that combines the fields of Robotics, Internet of Things and Cloud Robotics. The implantation of an IoT platform enabled communication between connected devices, provides the NAO robot with many additional capabilities, as well as indirectly linking the robot to the implemented HW node allows it to remotely use RESTful services. Connecting the REMEDES system to the IoT platform allowed the rest of the system entities to start the exercise and get the results remotely. Finally, creating an activity reminder application enabled us to exploit the capabilities of our system by interconnecting all the available "Things". As future work, the developed system can be expanded by enriching it with additional devices, sensors and actuators, as well as test its acceptance with real elders.